\documentclass{article}
\usepackage[a4paper, margin=2.5cm]{geometry}
\usepackage{hyperref}
\usepackage{amsmath, amssymb, amsfonts, natbib}
\usepackage{graphicx}
\usepackage{epstopdf}
\usepackage{epsfig}
\title{Social Mobility in India}
\author{
A. Singh,
Department of Economics and Finance, Pilani Campus, India,\\
email: \href{maito: anu.2singh7@gmail.com}{anu.2singh7@gmail.com}
\and
A. Forcina,
Dipartimento di Economia, Perugia, Italy
\and
K. Muniyoor,  	
Department of Economics and Finance, Pilani Campus, India
}
\begin{document}
\maketitle
\begin{abstract}
Rapid rise in income inequality in India is a serious concern. While the emphasis is on inclusive growth, it seems difficult to tackle the problem without looking at the intricacies of the problem. Social mobility is one such important tool which helps in reaching the cause of the problem and focuses on bringing long term equality in the country. The purpose of this study is to examine the role of social background and education attainment in generating occupation mobility in the country. By applying an extended version of the RC association model to 68th round (2011-12) of the Employment and Unemployment Survey by the National Sample Survey Office of India, we found that the role of education is not important in generating occupation mobility in India, while social background plays a critical role in determining one's occupation. This study successfully highlights the strong intergenerational occupation immobility in the country and also the need to focus on education. In this regard, further studies are needed to uncover other crucial factors limiting the growth of individuals in the country.
\end{abstract}
\begin{center}
\end{center}
\section{Introduction}
The notion of social mobility is related to equality of opportunities so that individuals can achieve higher social position regardless of the social background of their parents. It has two motivations, first, by allowing a better utilization of available talents it leads to increased overall efficiency and productivity in the labour market; second, its objective seems more realistic than equality of outcomes among citizens, which is a desirable objective under many point of view, \citep{corak2020intergenerational}. It encourages human capital investment that can be made equally available to all sections of society through better public institutions and its policies. While equality of opportunities leads to more social mobility, higher income inequality threatens social mobility. In this context, the famous Great Gatsby Curve shows negative cross-country relationship between income inequality and inter-generational mobility mentioned in \cite{corak2013income}; which suggests that inequality skews opportunity and lowers inter-generational mobility.

In ancient India, education, skills and occupation were determined by the caste of a person, thus there was not much freedom for moving between different levels of society \citep{deshpande2010history}. Although, since 1950, the emphasis was on abolishing the caste structure and providing equal opportunities to all, but strong limitations still exist in the country's occupational structure as shown by \cite{reddy2015changes}. Within the same period, the country has experienced a substantial increase in income inequality which can be proved by the fact that the share of the top ten percent income group in national income is increasing and the share of middle 40 percent and lower 50 percent income groups is decreasing \citep{chancel2019indian}. Interestingly, during this period, the country has also experienced rapid economic growth. In this regard, \cite{aiyar2020inequality} concludes that the low level of inter-generational mobility may be the cause why high economic growth coexists with rising income inequality.

If we consider this to be a meaningful explanation in the case of India, then it would be interesting to study more in depth of occupational intergenerational mobility to get a better understanding about the current situation in the country. As education is considered to be directly associated with occupation, if we take education and occupation together, it is possible to realize whether education supports occupational mobility. If it is not supported with the attainment of education, then a conclusion can be drawn about the direct transmission of occupation which mostly goes against the idea of equality of opportunities. Thus, social mobility in this study includes measurement of occupational inter-generational mobility as in \cite{erikson2002intergenerational} and social background is measured by the occupation of the individual's parent.

In this regard, the present paper attempts to examine the three principal research questions in the area of social mobility which are: (i) Do mostly sons of fathers with high level of occupation get higher education? (ii) Do mostly sons with higher education enter a higher level of occupation? (iii) on the whole, how strong is the association between the occupation of fathers and sons ? The purpose of this paper is to look at the current occupational immobility by associating it with educational attainment and social background. So, this investigation is based on the assumption that occupational mobility depends on educational attainment and social background. We use 68th round of NSSO data for our study which has been extensively used to study intergenerational mobility. By using an extended version of the Row-Column (RC) association models, which has never been applied before within the mobility field, we expect to complement the existing literature.

We find that the association of an individual's social background with his education is moderate, while this relationship is quite strong with occupation. This is because, according to our results, education does not seems to play a huge role in deciding one's occupation in India. These findings are consistent with existing literature and it emphasizes the lack of quality education in the country. The rest of the paper is organized as follows: section 2 reviews the existing literature on inter-generational education and occupational mobility, section 3 deals with description of the data and socio-economic characteristics of the working sample, section 4 discusses association method and section 5 presents results and its analysis followed by the discussion and conclusions in section 6.

\section{Studies on inter-generational occupation mobility and education attainment}
The first area of study relevant for this paper is human capital theory which was developed by Becker and Mincer and focuses on parents' decision to invest in children's education and its impact on their income and occupation levels \cite{becker1979equilibrium}. Parents investing on the education of their children may be seen as a way to affect the occupation they may obtain by investing to provide them with better skills and knowledge.

Status attainment theory focuses on additional factors, above and beyond the level of schooling, by which parents transfer, by family interactions, life styles and other advantages to their children that persist throughout life, including prospective adult wage advantages \citep{haveman1995determinants}. It may work by direct transfer of benefits from parents to their children if, for example, the son of a father with a better profession may get the same occupation due to family ties.

Next, Weber's concept of social closure discusses how "social collectives seek to achieve maximum rewards by limiting access to resources and opportunities to a limited circle of eligible" \citep{parkin1979marxism}. For example, in order to get admission in good universities, if a person needs certain qualities, which are generally available among children from affluent backgrounds, then it will prove to be an obstacle for children with less fortunate background to get admission in such universities \citep{fishkin2012unequal}.

Concerning empirical studies, we now review some applications with reference to the situation in India. Using National Election Study (NES) data of 1996, \cite{kumar2002determinants} described occupation mobility in terms of origin and destination. They found that 90 percent of the people in farming came from farming background which may be due to transfer of land from father to his son. Salary class (which usually consists of white collar and skilled occupations), apparently reach their position starting from fathers of diverse backgrounds. Also, 68 percent individuals from unskilled background remain unskilled.

Along the same line, \cite{motiram2012close} using the first round of India Human Development Survey of India (IHDS-1), showed that mostly the sons of unskilled and low paid fathers remain in the same occupation.
Another study on education and occupation inter-generational mobility using National Sample Survey Office (NSSO) rounds from 1983 to 2005 has shown convergence in rates of conditional probabilities of education mobility among non SC/STs and SC/STs caste groups \citep{hnatkovska2013breaking}, which suggests that differences in rates of mobility between these two groups have reduced, however, when it comes to occupational mobility, stagnation still exists which is due to factors other than caste. The Scheduled Castes (SCs) and Scheduled Tribes (STs) are among the most disadvantaged socio-economic groups in India. \cite{hnatkovska2013breaking} used median wages to classify occupations, and EUS data usually has many missing values in wages and incomes, mainly for self-employed farmers whose proportion is large in rural India. Next, they kept grandfather and father in the same generation and the child and the grand child together in the next generation which is usually not appropriate when we want to explore the mobility between adjacent generations. Further, they used regression and transition matrices to measure the education and occupation mobility. The probit regression, on the one hand, does not take into account the distance between the occupations of the father and son and only observes whether the son leaves the father occupation and the transition matrix only shows the distribution pattern.

\cite{reddy2015changes} measures changes in the occupational mobility using the same data up to the year 2011-12. In this, the author suggested, there exists less occupational inter-generational mobility in India, especially among the Scheduled Castes (SCs) and Scheduled Tribes (STs). We note that the method used in the above study is complex, involving few steps that can be avoided if using log linear or related interactions which are not affected by changes in the marginal distributions. It is useful to mention that the interaction parameters in the RC model are not affected by the marginal distribution, so there is no need of standardizing the mobility tables required to have the same occupational distributions as in \cite{reddy2015changes}.

With regard to education mobility, \cite{kishan2018past}, by looking at the correlation between father and sons' years of schooling, suggest education mobility. On the same line, \cite{ray2010educational}, using the 1993 and 2004 NSSO rounds, suggested less mobility for both occupation and education, with occupational mobility being less than education mobility. Next, \cite{azam2015like}, using the first round of the IHDS data, estimated average inter-generational correlation for India at 0.523 which is higher than the average global correlation of 0.420. Also, they suggested strong association between expenditure on education with the estimated inter-generational mobility in education attainment.

\cite{mueller2000structure} compared association between occupation and education mobility between the United States and Germany using the International Social Survey of Program (ISSP) 1987 for Germany and General Social Survey 1994 for the US. The author finds that social origin have a strong ties with education attainment which is associated with later access to occupation opportunities. For instance, higher education has strong ties with white-collar occupations. In comparison, Germany has been shown to have more mobility than the United States. \cite{meyer1979education}, compared occupation and education mobility between Polish men and American men using regression analysis on the 1972 and 1976 survey data sets. They also suggested that the type of school determines occupational attainment. Further, \cite{carnevale2011college} used the American Community Survey 2007-09 to predict higher education opens up access to higher paid jobs through the use of synthetic estimates of work life earnings. Finally, we were unable to find much studies on the association of education with occupation  mobility in the Indian context. In addition, the use of RC models has been more recent in this area through the use of mobility tables, which we expect will strengthen the existing literature.
\section{Description of the data}
The data used in this paper come from the 68th round (2011-12) of the Employment and Unemployment Survey (EUS) conducted by the National Sample Survey Office (NSSO) of India. The EUS provides primary source of data for various indicators of labour force at state and national level. It follows a stratified multi-stage sample design and includes a sample of around 100,000 households covering almost all geographical regions of the country. It is the largest data gathering information on almost every social and economic aspect at the individual and household level since 1983 in India. It contains information about education in 13 broad categories ranging from not literate to graduate and above and occupation levels are classified according to the national classification of occupations (NCO-2004) four- digit occupation codes. The basis of divisions in the occupational structure is based on the skills required to perform the functions and duties of an occupation.

Initially we arranged the education categories into six groups: not literate, without formal schooling, primary, secondary, higher secondary or diploma certificate and graduate and above that ranged from 1 to 6, respectively. However, because the proportion of sons in the second category of education is less than 0.2 percent in our sample, we decided to merge categories 1 and 2, thus, in the analysis, education is taken as having 5 categories. We categorized occupation codes into four categories as unskilled, farming, skilled/semi-skilled and white collar respectively by following the NCO single-digit occupation codes \cite{MinofLabor2004NCO} and \cite{reddy2015changes} occupational structure.
It is worth noting that there is no uniformity in selecting the framework of occupational structure as literature exists with different structural frameworks by different authors in the context of the same country.
Here, the unskilled occupation includes labours from agriculture and fisheries, mining and construction activities. The farming business includes market oriented skilled and subsistence agriculture and fishery workers. Skilled and semi-skilled occupations include office clerks, service workers, sales workers, craft-related trades workers, plant and machine operators, and assemblers. White collar occupations include legislators, managers, professionals.

The NSSO data does not contain information about parents if the person is living separately from his family. Therefore, in order to do study on inter-generational mobility, we selected only those households where the working person and his father are living together. Also, we concentrate on male subjects because married women in India live with their husbands or father-in-law and the survey does not provide information on their parents. Thus, the criteria for selecting the working sample were households where the son's age was between 16 and 45 and both father and son were not currently enrolled in any educational institution and informed about their education and occupation. The above criteria for sample selection provides a sample of working father and son from which we removed cases where the required information was missing. This procedure lead to a sample of 27771 households which is our 'working sample'. In case a father was living with more than one working age son, we selected only the eldest son to ensure that we are obtaining the record of a father and a son in our working sample.

To check whether the selection leading to our working sample is unbiased, we compared the socio-economic characteristics of co-resident sons with sons who are living separately from their fathers. In practice, non co-resident sons correspond to households with only one adult male who is of working age. We found 48390 non-co-resident households in our sample.

\begin{table}[htb]
\caption{\label{tab:1} Summary statistics for sons who are co-resident or are living on their own}
\centering
\fbox{
\begin{tabular}{lcrrrcrrr} \\ \hline
 & \vspace{1mm} & \multicolumn{3}{c}{co-resident} & \vspace{1mm} &
 \multicolumn{3}{c}{Living on their own} \\
Variable & & Obs & Mean & SD & & Obs & Mean & SD  \\ \hline
Age              & & 27771 & 25.91 & 6.12    & & 48390   & 35.83 &   6.54\\
\% of Rural Pop. & & $\dots$ & 69.72       & & $\dots$ &      57.87 &   \\
\% of SC/ST      & & $\dots$ & 27.24       & & $\dots$ & 35.83 & 6.54 \\
Years of Education & & $\dots$ & 9.94 & 3.41     & & $\dots$ & 8.84 & 4.46 \\
Log MPCE         & & $\dots$ & 7.14 & 0.54 & & $\dots$ & 7.26 & 0.59 \\
\\ \hline
\end{tabular}}
\end{table}
\begin{table}[htb]
\caption{\label{tab:2} Occupational distribution of sons by living arrangement}
\centering
\fbox{
\begin{tabular}{llcrr} \\ \hline
Description & Score & \vspace{1mm} & Co-resid. & On their own \\
 \hline
Unskilled	(U) & 1 & & 17.27 &	17.55 \\
Farming		(F) & 2 & & 29.82 &	16.66 \\
Skilled/Semi (S) & 3 & & 36.19 & 43.35 \\
White collar (W) & 4 & & 16.73	& 22.44
\\ \hline
\end{tabular}}
\end{table}
\begin{table}[htb]
\caption{\label{tab:3} Educational attainment of sons by living arrangement}
\centering
\fbox{
\begin{tabular}{llllrrr} \\ \hline
Description & Score & & Co-resid. & & On their own \\
 \hline
Without Schooling (N) & 1 & & 6.86 & & 16.80 \\
Primary (P) & 2 & & 19.08 & & 22.38 \\
Secondary (S) & 3 & & 43.86	& & 34.87\\
HSC/Diploma/Certificate (H) & 4 & & 16.81 & & 11.91 \\
Graduation and above (G) & 5 & & 13.39 & & 14.04
\\ \hline
\end{tabular}}
\end{table}

In addition we compared the frequency distributions relative to occupation of co-resident sons and non-co-resident sons. From table 1 we can see that except for some difference in the age between both the groups; years of education and log of monthly per capita expenditure (MPCE) is not significantly different. If we look at the distribution of occupation of co-resident sons and non-co-resident sons in Table 2, while the proportion of unskilled workers is similar between the two groups, farming occupation is more prevalent among co-residents. And skilled/semi and white collar occupations are more prevalent among individuals living separately from their parents. This is due to the fact that a large proportion of co-resident families exist in rural areas and hence the proportion of co-resident sons engaged in agriculture is higher than that of individuals living on their own. If we compare the education levels of co-resident sons with sons living on their own, then we then find that the proportion of sons up to primary level education is higher in the case of sons living separately, while co-resident sons have more persons with higher secondary and equivalent education. However, this gap was bridged between the two groups by individuals with the same level of graduate and above education. Thus, we believe that our working sample involving only co-resident households is representative and comparable, at least for the purpose of this study.

Descriptive statistics of our working sample show that the mean age of sons is 26 and father is 55. There are 6.86 percent sons without education while the father's generation comprises 33.66 percent people without education. The sons' generation consists of 43.86 percent of the people with secondary education, while people with graduate and above education is only 13.39 percent. However, it is better than the percentage of graduates and above in the fathers' generation, which is only 5.49 percent. Therefore, it is possible to say that the level of education has increased in the generation of sons, which is proved by the education of 9.93 average years in the sons' generation, where earlier it was only 6.42 average years in the father's generation. If we look at the level of occupation, then, the sons' generation is governed by skilled and semi-skilled occupations, which is 36 per cent and only 16.73 per cent white collar occupation. While, father's generation comprises mostly of farming occupation which is 39 percent and interestingly, no change has been recorded in the proportion of white collar occupation which is 17 percent in father's generation also.

\section{Statistical methods and social mobility}
Statistical methods suitable for the analysis of social mobility depend both on the nature of the data and on the purpose of the analysis. For instance, when, like in \cite{mazumder2016estimating}, one has income data at the individual level for the father and the son, methods based on linear regression on incomes or on the corresponding ranks may be used, depending on whether one believes that the relation is approximately linear or not. Instead, when, like in our case, data are in the form of contingency tables, methods based on interactions are more suitable. Another important distinction is whether one aims to summarize the overall degree of association by a single number like in \cite{altham2007comparing} or to undertake a more analytical investigation, looking at several measures of association at the same time.

There is substantial agreement in the literature that the set of log-linear interactions computed on a contingency table provide one of the best assessment of the strength and the direction of association between the row and column variable. Clearly, stronger association means that the social class of the son may be more easily predicted from that of the father, thus, stronger association is equivalent to smaller chances of social mobility. An important property of interaction parameters is that they are not affected by the structure of marginal distribution. This is related to the algorithm described in \cite{altham2007comparing} which allows to transform a given contingency table into another having the same set of interactions and arbitrary  marginal distributions. This may be important in the light of separating structural from relative or circulation mobility as discussed, for instance by \cite{hauser1988cross} and \cite{sobel1985exchange}.

It is well known that in an $r\times c$ contingency table, we can compute $(r-1)(c-1)$ non redundant log-linear interactions measuring the degree of immobility within different subsections of the table. There are, essentially, two different strategies to deal with such a multitude of measure: (i) to compute a unique summary measure by some appropriate average as in \cite{altham2007comparing}, an approach applied, for instance, in \cite{reddy2015changes}, or (ii) try to fit some restricted model depending on a smaller number of parameters, a route followed in this paper where RC association models are applied. RC association models were introduced by \cite{Goodman81} to simplify the association structure without loosing important information.
These models have been used for the analysis of social mobility by, for instance, \cite{xie1992log} and \cite{mueller2000structure}. An RC(1) association model has just one coefficient of intrinsic association: higher values of this coefficient indicate stronger association and thus lower mobility. In addition, the estimated model provides a set of row and column scores from which we can measure the relative distance between categories: if two categories are close to each other, the corresponding conditional distributions are very similar.

Various extensions of log-linear interactions have been studied in order to capture more specific features of association; they are essentially based on assigning a logit of type L (local), G (global) or C (continuation) to the row and the column variables.  A wide collection of interaction parameters obtained by combining different row and column logit types are studied in \cite{douglas1990positive} in the context of positive association, a notion closely related to social mobility when  father and son social class may be ordered from lowest to highest, in that case, stronger positive association means lower mobility.
\cite{douglas1990positive} also provide a graphical interpretation of the different interaction parameters.
RC association models may be used to extract the most relevant features of the association structure in a social mobility table when interactions are defined by combining row and column logit types, see for instance \cite{Bart02}. One further extension, introduced by \cite{Kateri95}, has allowed to combine traditional RC association models, Correspondence analysis and a whole collection of other models into a unified class of RC association models depending on a scaling factor.

The statistical methods used in this paper are based on the even larger class of RC association models of \cite{Forcina20} which allow the user to choose both the type of interaction parameters as in \cite{douglas1990positive} and the scaling factor as in \cite{Kateri95}. The advantage of this approach is that we may easily explore a large range of different models and select the one that is as simple as possible and fits the data best. The strategy used in this paper is to search for the smallest $K$ such that an $RC(K)$ model fits the data sufficiently well. For the three tables analysed in this paper, no satisfactory model with $K=1$ seemed to be adequate; on the other hand, it was possible to find an $RC(2)$ model which fits the data very accurately.
While the deviance is uniquely defined, computations of the coefficients of intrinsic associations and the rows and columns scores depend on row and column weights; we adopted the usual strategy \citep[see][Chap. 6]{Kateri14} based on uniform weights.

The strength of immobility in an RC(2) model depends on two coefficients of intrinsic association, where higher association means more immobility. To give an idea of the degree of immobility implied by a given pair of coefficients, below we compare several hypothetical version of the association between father occupation and son education. More precisely, we consider the joint frequencies that we had got if, keeping the rows and columns score fixed to the vales estimated by the best model, the pair of coefficients of intrinsic association, relative to the values estimated in the best fitted model were: a - the same, b - both divided by two, c - both multiplied by 2.5.
\begin{table}[htb]
\caption{\label{tab:4} Theoretical joint frequencies for the education of sons of fathers in U and W in three hypothetical scenarios}
\centering
\fbox{
\begin{tabular}{lcrrrrr}
 & \vspace{1mm} & \multicolumn{5}{c}{Son Education} \\
Father & & N & P & S & H & G \\ \hline
U, a & &  525  &  1221  &   1711  &    304  &   103\\
U, b & &  381  &   898  &   1639  &    550  &   396\\
U, c & & 1325  &  1690  &    756  &     77  &    16\\ \hline
W, a & &  134  &   499  &   1691  &   1073  &  1377\\
W, b & &  207  &   707  &   2108  &    943  &   809\\
W, c & &    0  &    13  &   1047  &   1609  &  2106\\
\hline
\end{tabular}
}
\end{table}

\section{Social mobility in India}
\subsection{Father occupation and son education}
We now study the joint distribution of father's occupation and son's educational attainment in India. This will help us understand to what extent educational attainments of the son depends on his father's occupation in the sense that father with a better occupation have better chances to invest more in the education of their sons.

\begin{table}[htb]
\caption{\label{tab:5} Observed joint distribution of households by father occupation and son education}
\centering
\fbox{
\begin{tabular}{lcrrrrr}
 & \vspace{1mm} & \multicolumn{5}{c}{Son Education} \\
Father Occ & & N & P & S & H & G \\ \hline
U & &  526   &   1222  &   1707  &    307  &   102\\
F & & 731    &  1911   &    5046  &     1916  &   1224\\
S & &   512  &   1664  &   3742  &   1370   &  1017\\
W & &  135    &   501  &   1686  &    1076  &   1376\\
\hline
\end{tabular}
}
\end{table}

\begin{figure}[htb!]
\centering
\includegraphics[width=0.9\textwidth]{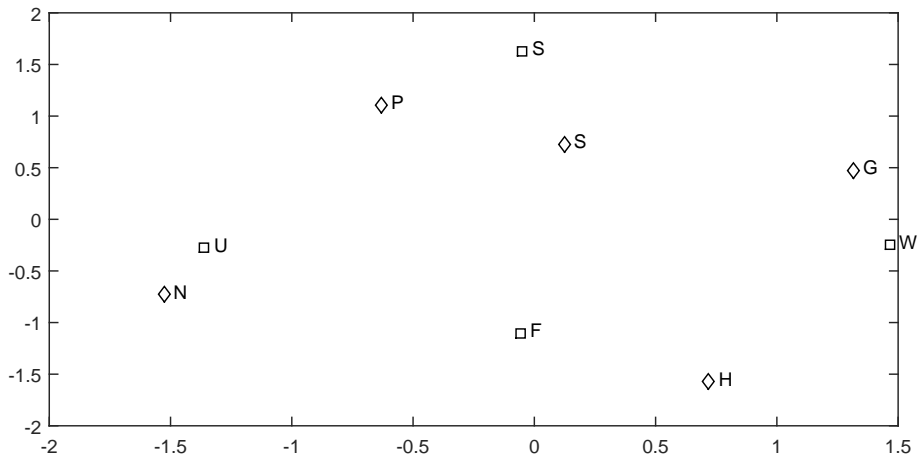}
\caption{\label{Fig:1} \it Plot of row (squares) and column (diamonds) scores for the data in Table \ref{tab:5}}.
\end{figure}

At first, a collection of extended RC(1) models as in \cite{Forcina20} were fitted by setting logit type for occupation to L because its categories are not necessarily ordered and L, G and C for education, for a range of values of the $\lambda$ parameter; the best of these models had deviance of about 28 on 6 degrees of freedom, which is significant. Thus, we moved to RC(2) models: the best fit was obtained by setting logits to L for occupation and G for education with $\lambda=0.22$. This model has a deviance of 0.14 on 2 degrees of freedom. The coefficients of intrinsic association are equal to 0.99 and 0.02 respectively. The row and column scores are plotted in Figure \ref{Fig:1}

It is interesting to note that on the horizontal axis, which corresponds to the largest coefficient, education categories are ordered and almost equally spaced; the same is true for occupation, but F and S are very close to each other on the horizontal axis. Note, also that the pairs (U,N) and (W,G) are both close suggesting that the sons of unskilled workers are the most likely to achieve no formal education while those of white collars are the most likely to get a G+ degree. However, from the frequency distribution in table 5, we see that about 75\% of the sons of unskilled father get primary or secondary education. Probably, this is the result of schemes like Sarva Shiksha Abhiyan (SSA), Mid-Day Meal Scheme, Right to Education (RTE) Act which have helped children from poor backgrounds get enrolment up to secondary level. On the horizontal axis S (education) is very close to both F and S, indicating that a large proportion of sons of farmers or skilled-semi skilled workers get secondary education.
On the whole, considering also the coefficients of intrinsic association, we may say that the effect of father occupation on son education is active but to a moderate degree, allowing for a reasonable amount of mobility. However, much remains to be done to enable and cover such students in the workforce.
\subsection{Son education and son occupation}
The purpose of the following analysis is to determine how much the the efforts spent in getting a better education improve the chances of getting a better job, in other words we examine the role of education in achieving higher level jobs in India. It is worth noting that here strong association means, roughly, that people get the job for which they are qualified, instead, weak association indicates that other factors, like family influence and connections, play an important role.
\begin{table} [htb]
\caption{\label{tab:6}Joint distribution of households by son education and son occupation}
\centering
\fbox{
\begin{tabular}{lcrrrr}
 & \vspace{1mm} & \multicolumn{4}{c}{Son occupation}  \\
Son Education & & U & F & S & W \\\hline
N   & &     655    &     579   &      572     &     98\\
P   & &    1544   &      1461   &      1933    &     360\\
S   & &   2134   &     3996    &     4739    &     1312\\
H   & &    363    &      1466   &      1749    &     1091\\
G   & &     100   &       778    &      1057     &     1784
\\ \hline
\end{tabular}
}
\end{table}

\begin{figure}[htb!]
\centering
\includegraphics[width=0.9\textwidth]{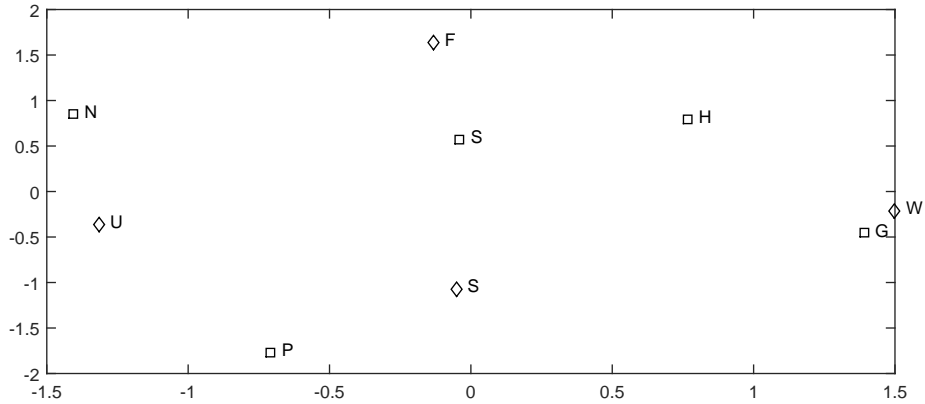}
\caption{\label{Fig:2} \it Plot of row (squares) and column (diamonds) scores for the data in Table \ref{tab:6}}.
\end{figure}

Some preliminary model selection suggested that no RC(1) model fits sufficiently well the data, so we examined a range of RC(2) models, the one with logit type C for education and L for occupation with $\lambda$ = -0.06 fits best with a deviance of 0.11, which means an almost perfect fit. The estimated coefficients of intrinsic association are equal to 1.03 and 0.02 respectively. The row and column scores are plotted in Figure \ref{Fig:2}

The plots in Figure \ref{Fig:2} indicate, again, that, on the horizontal axis, education categories follow the natural order and are almost equally spaced. Occupational categories follow also the expected order on the horizontal axis, except that F and S are very close to each other, though they diverge on opposite directions on the vertical axis. The fact  that G and W are very close to each other, means that sons with a G+ degree have the highest chances of becoming white collars. The same is true for the pair N and U, but only on the horizontal axis: the most likely occupation for sons with N (education) is U, but they have non negligible chances of ending up into F or S. On the horizontal axis, S (education) is between F and S (occupations), meaning that sons with secondary degree are most likely to become farmers or skilled workers. On the whole, the strength of association is only a little stronger than in the previous table, meaning that education is not the only factor that determines the kind of occupation that a person can acquire.
\subsection{Father occupation and son occupation}
The purpose of the following analysis is to examine the shape and strength of association between father occupation and son occupation. This is important to answer the following question: the effect of father's occupation on son's occupation is only indirect, that is induced by the fact that fathers with a better occupation can afford to invest more to provide a better education to their sons who, because of their education, can get a better job, or there is also a direct effect, in the sense that the sons of fathers with a better occupation, because of family ties, can get a similar occupation even if not adequately qualified.
\begin{table} [htb]
\caption{\label{tab:7} Joint distribution of households by son education and son occupation}
\centering
\fbox{
\begin{tabular}{lcrrrr}
 & \vspace{1mm} & \multicolumn{4}{c}{Son occupation} \\
Father occupation & & U & F & S  & W  \\\hline
U  & &  2644   &       192   &      898   &      130\\
F  & &   988  &      6861   &      1917   &     1062\\
S  & &   845   &       730   &     5952   &     778\\
W  & &   319   &      497   &      1283    &    2675 \\ \hline
\end{tabular}
}
\end{table}
For these data all RC(1) models fit badly irrespective of the logit types while the RC(2) fits very well, so we set both logit types to G and searched for the optimal value of $\lambda$ which equals -1.21 with a deviance of 0.01 on 1 degree of freedom. The two coefficients of intrinsic association equal to 3.35 and 0.24 respectively, almost three times larger than in the previous two cases above, indicating that, probably, family ties must be operating in addition to education.

Both the rows and columns scores follow the natural order on the horizontal axis which is the most important. Note also that each category of father occupation is fairly close to the corresponding category of son occupation, at least on the horizontal axis, which suggests that, to some degree, sons tend to remain in the same occupation of their father; indeed, the largest frequencies are along the main diagonal in Table \ref{tab:7}
\begin{figure}[htb!]
\centering
\includegraphics[width=0.9\textwidth]{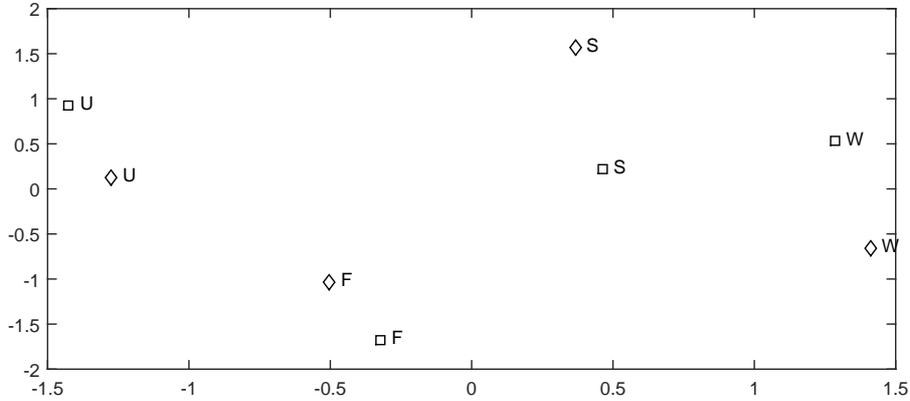}
\caption{\label{Fig:3} \it Plot of row (square) and column (diamonds) scores for association between father and son occupations in Table \ref{tab:7}}.
\end{figure}

The above analysis shows that the association of father occupation to son occupation is strong. This implies that regardless of a person's education background, a son is more likely to get the same occupation of his father. Thus, it can be concluded that the connection is direct rather than mediated through education. If we try to match the ground reality with our results, then our results match the practical aspect prevailing in India. In India, it is found to a large extent that the father tries to keep his child in his profession. This may be due to less return from education as in \citet{Shrivastava2019,Aggarwal2014} and hence father's influence in the labour market predominates in deciding his child's profession. This is consistent with the inference that wherever there is less return from education and skills, occupation pathway becomes the primary channel of inter-generational persistence \citep{blanden2014intergenerational}.
\section{Conclusion}
% O  F1        S1        F4        S4
%    0.1415    0.2119    0.0753    0.1034
%    0.5086    0.4676    0.3896    0.2928
%    0.2039    0.2162    0.2922    0.3674
%    0.1460    0.1042    0.2429    0.2364
% E  F1        S1        F4        S4
%    0.3814    0.0845    0.2447    0.0464
%    0.3310    0.2400    0.2609    0.1431
%    0.1927    0.4502    0.3192    0.4019
%    0.0475    0.1389    0.0775    0.2031
%    0.0473    0.0865    0.0978    0.2055

In this paper we have investigated inter-generation social mobility in India by using the 68th round of NSSO data for 2011-12 year. Our results indicate that the association between father occupation and son educational attainments is moderate, meaning that, probably because of the present policies of the government, together family efforts, the sons coming from a modest background have over 50\% chances to reach, at least, secondary education. Unfortunately, the association between son education and son occupation is also moderate, indicating that education is not the main factor that determines occupation and, thus, social position. This finding is confirmed by the fact that the association between father and son occupation is much stronger than those passing through education. This means that there are other factors that determine one's occupation apart from education. Overall, it suggests that the role of social background in deciding one's education is only moderate while the role of the same social background is strong for deciding one's occupation. The strong dependence of occupation on social background suggests that India is still not an open society and especially opportunities for work are not quite distributed.

We believe that there are three important interpretations for the above paradigms of social mobility in India. First, India's social structure evolved from a rigid caste structure but still there exist restrictions in society especially at the lower level, which do not allow certain groups to grow and take advantage of development. Second, the limited role of education in determining one's occupation also exists due to unsatisfactory quality of education in the country. This is proved by the fact that, despite several initiatives taken by the government at the lower level of education, only 9 out of 28 states have shown improvement in the School Education Quality Index (SEQI, 2019), while for 9 states it has gone down and the rest show no change as per National Institution for Transforming India (NITI Ayog). Further, if we look at India's position in advanced education, its score is 56.42 which is one of India's lowest component scores in the Social Progress Index (SPI 2020). At the same time, if we look at the component score for the quality of education of Scandinavian countries, it is quite higher than many countries in the world. Overall, their ranking in the Global Social Mobility Index 2020 and SPI 2020 is quite high and the rate of inequality is also very low in these countries. Thus, it is possible to say that social mobility, which has been seen as an important tool to bring long term equality, has a clear link with fair education and occupational opportunities in the country.
Third, other important factors such as health, infrastructure and technology are currently under development in the country, which directly contribute to the above social mobility indicators. Since India's resources are diverse and the requirements of one state may be different from others, a state-level study on social mobility indicators at the national level will help identify the lack of components at the national level and demonstrate the need for immediate improvement at the regional level. We intend to study social mobility indicators at the state level in subsequent work.

\end{document}